\documentclass[pdflatex,sn-mathphys,iicol]{sn-jnl}% Math and Physical Sciences Reference Style
\jyear{2023}%

\theoremstyle{thmstyleone}%
\theoremstyle{thmstyletwo}%

\theoremstyle{thmstylethree}%

\usepackage{etoolbox}   
\makeatletter
\patchcmd{\@maketitle}{\artauthors}{{\artauthors}}{}{}
\makeatother

\usepackage{ulem}
\newcommand{\referee}[2]{\textcolor{orange}{#2}}

\begin{document}

\title[$ $]
{\vspace{-2cm}
The photometric variability of massive stars due to gravity waves excited by core convection}
\graphicspath{{./}{figures/}}


\author*[1]{\fnm{Evan H.} \sur{Anders}}\email{evan.anders@northwestern.edu}
\author[1,2]{\fnm{Daniel} \sur{Lecoanet}}
\author[3,4]{\fnm{Matteo} \sur{Cantiello}}
\author[3,5]{\fnm{Keaton J.} \sur{Burns}}
\author[1,2]{\fnm{Benjamin A.} \sur{Hyatt}}
\author[1,6]{\fnm{Emma} \sur{Kaufman}}
\author[7]{\fnm{Richard H.~D.} \sur{Townsend}}
\author[8]{\fnm{Benjamin P.} \sur{Brown}}
\author[9]{\fnm{Geoffrey M.} \sur{Vasil}}
\author[10]{\fnm{Jeffrey S.} \sur{Oishi}}
\author[3]{\fnm{Adam S.} \sur{Jermyn}}

\affil[1]{\orgdiv{Center for Interdisciplinary Exploration and Research in Astrophysics (CIERA)}, \orgname{Northwestern University}, \orgaddress{\street{1800 Sherman Ave}, \city{Evanston}, \postcode{60201}, \state{IL}, \country{USA}}}

\affil[2]{\orgdiv{Department of Engineering Sciences \& Applied Mathematics}, \orgname{Northwestern University}, \orgaddress{\street{2145 Sheridan Road}, \city{Evanston}, \postcode{60208}, \state{IL}, \country{USA}}}

\affil[3]{\orgdiv{Center for Computational Astrophysics}, \orgname{Flatiron Institute}, \orgaddress{\street{160 Fifth Avenue}, \city{New York}, \postcode{10010}, \state{New York}, \country{USA}}}

\affil[4]{\orgdiv{Department of Astrophysical Sciences}, \orgname{Princeton University}, \orgaddress{\street{4 Ivy Lane}, \city{Princeton}, \postcode{08544}, \state{NJ}, \country{USA}}}

\affil[5]{\orgdiv{Department of Mathematics}, \orgname{Massachusetts Institute of Technology}, \orgaddress{\street{77 Massachusetts Avenue}, \city{Cambridge}, \postcode{02139}, \state{MA}, \country{USA}}}

\affil[6]{\orgdiv{Department of Physics \& Astronomy}, \orgname{Northwestern University}, \orgaddress{\street{2145 Sheridan Road}, \city{Evanston}, \postcode{60208}, \state{IL}, \country{USA}}}

\affil[7]{\orgdiv{Department of Astronomy}, \orgname{University of Wisconsin-Madison}, \orgaddress{\street{475 N. Charter Street}, \city{Madison}, \postcode{53706}, \state{WI}, \country{USA}}}

\affil[8]{\orgdiv{Department of Astrophysical \& Planetary Sciences}, \orgname{University of Colorado Boulder}, \orgaddress{\street{391 UCB}, \city{Boulder}, \postcode{80309}, \state{CO}, \country{USA}}}

\affil[9]{\orgdiv{School of Mathematics}, \orgname{University of Edinburgh}, \orgaddress{\street{James Clerk Maxwell Building}, \city{Edinburgh}, \postcode{EH9 3FD}, \country{UK}}}

\affil[10]{\orgdiv{Department of Physics \& Astronomy}, \orgname{Bates College}, \orgaddress{\street{2 Andrews Road}, \city{Lewiston}, \postcode{04240}, \state{ME}, \country{USA}}}

\abstract{

Massive stars die in catastrophic explosions, which seed the interstellar medium with heavy elements and produce neutron stars and black holes. 
Predictions of the explosion’s character and the remnant mass depend on models of the star's evolutionary history. 
Models of massive star interiors can be empirically constrained by asteroseismic observations of gravity wave oscillations. 
Recent photometric observations reveal a ubiquitous red noise signal on massive main sequence stars; a hypothesized source of this noise is gravity waves driven by core convection.
We present the first 3D simulations of massive star convection extending from the star's center to near its surface, with realistic stellar luminosities. Using these simulations, we make the first prediction of photometric variability due to convectively-driven gravity waves at the surfaces of massive stars, and find that gravity waves produce photometric variability of a lower amplitude and lower characteristic frequency than the observed red noise. We infer that the photometric signal of gravity waves excited by core convection is below the noise limit of current observations, so the red noise must be generated by an alternative process.
}

\pacs[UAT Keywords]{Stellar Photometry (1620), Internal waves (819), Stellar Oscillations (1617), Hydrodynamical Simulations (767), Stellar Convection Zones (301), Stellar Cores (1592)}

\maketitle

\section{Introduction}
\label{sec:intro}

The oxygen we breathe was generated in the cores of massive stars \citep{rauscher_etal_2002} and expelled into the interstellar medium in violent explosions \citep{woosley_etal_2002} before mixing with the molecular cloud that formed our solar system \citep{kennicutt_evans_2012}. 
In addition to producing the elements that enable life, massive stars leave behind compact remnants, whose subsequent mergers have provided a new window into the universe through gravitational wave astronomy \citep{abbott_etal_2017}.
Predictions of the elemental yield of a massive star and the nature of its remnant are sensitive to many processes including convection {\citep{kaiser_etal_2020}}, winds {\citep{smith_2014}}, nuclear reaction rates \citep{fields_etal_2018}, and numerical algorithms {\citep{farmer_etal_2019}}.
Empirical constraints upon the interior structures of massive stars {\citep{pedersen_etal_2021}} could reduce uncertainties in these processes, which would improve stellar evolution calculations, and in turn models of star formation {\citep{mckee_ostriker_2007}}, galaxy formation \citep{pillepich_etal_2018}, and the reionization of the early universe \citep{bromm_larson_2004}.

It was recently shown that photometric light curves of hot, massive stars contain a ubiquitous red noise signal \citep{bowman_etal_2019, bowman_etal_2019_nat, bowman_etal_2020, dornwallenstein_etal_2020, szewczuk_etal_2021, bowman_dornwallenstein_2022}.
Theories for the driving mechanism of red noise include gravity waves stochastically excited by core convection, or turbulence from subsurface convection zones \citep{bowman_etal_2019_nat,cantiello_etal_2021,schultz_etal_2022}. 
Of particular interest is the hypothesis that these signals are waves driven by core convection, which could empirically constrain interior models of massive stars.

We present here the first simulations of core convection that include all the relevant physics to make an accurate prediction of the stellar photometric variability from convectively-generated waves.
Our simulations build upon pioneering studies which examined the properties of the turbulent core convection {\citep[e.g.,][]{browning_etal_2004, gilet_etal_2013, augustson_etal_2016, higl_etal_2021, baraffe_etal_2023}}, the characteristics of the gravity waves in the radiative envelope \citep{rogers_etal_2013, rogers_mcelwaine_2017, couston_etal_2018,edelmann_etal_2019,horst_etal_2020,lecoanet_etal_2021,lesaux_etal_2022,blouin_etal_2023}, and the possible observational features of those waves \citep{lecoanet_etal_2021, samadi_etal_2010,shiode_etal_2013,aerts_rogers_2015,lecoanet_etal_2019,thompson_etal_2023}.

Previous studies focused on the \textit{shape} of the frequency spectrum of gravity waves from convection \citep{aerts_rogers_2015,edelmann_etal_2019,horst_etal_2020,thompson_etal_2023}, but could not constrain the \textit{amplitude} of the waves because their simulations do not extend to the surface and use boosted convective luminosities.
Here we present simulations with realistic stellar luminosities which we use to predict \textit{both} the amplitude and shape of the frequency spectrum of photometric variability due to gravity waves excited by core convection.

Predicting this frequency spectrum is challenging because of the range of length scales, time scales, and physical processes important in the star \citep{jermyn_etal_2022_atlas}.
Waves are generated on fast convective timescales ($\sim 14$ days), but as they propagate to the surface they excite resonances with much longer lifetimes ($\gg 4\, {\rm yrs}$ from Kepler data \citep{kurtz_etal_2015}; $\sim 10^5\,{\rm yrs}$ from theoretical estimates \citep{shiode_etal_2013}).
It is unfeasible to simultaneously resolve these timescales.
To determine the wave signal at the surface, we appeal to an acoustic analogy.

The character of music depends both on the sound waves produced by musicians and on the acoustics of the environment where it is played {\citep[e.g.,][]{pentcheva_abel_2017}}.
Music is recorded in special studios with walls that absorb or diffuse waves to minimize the influence of the environment on the sound and retrieve the ``pure sound'' of the musicians.
To experience music in a different environment, it is not necessary to physically transport the musicians; instead, one can apply a filter to the recording, mimicking the effects of the new environment.

\begin{figure*}[ht!]
    \centering
    \includegraphics[width=\textwidth]{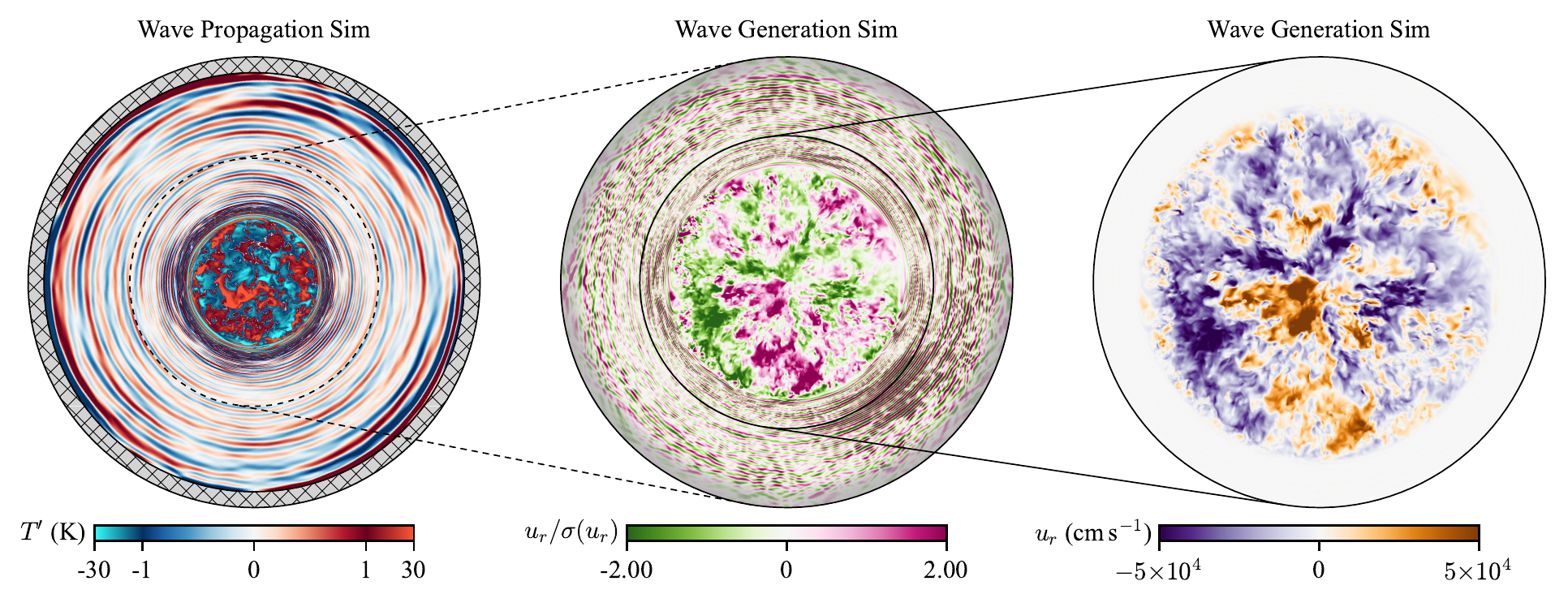}
    \caption{
    %Equatorial slices of instantaneous dynamics in simulations 
    Snapshots of equatorial slices through 3D spherical simulations of our fiducial 15 $M_{\odot}$ star.
    (Left Panel) Temperature fluctuations around the mean profile are shown in our ``Wave Propagation'' simulation; this simulation spans 93\% $R_*$, and the hatched region displays the $0.07 R_*$ that we do not simulate.
    The gravity waves cause perturbations of $\pm 1$ K, while the core convective temperature perturbations are of order $\pm 30$ K.
    (Middle Panel) The radial velocity field in our ``Wave Generation'' simulation is shown, normalized by the standard deviation of the radial velocity at each radius.
    The Wave Generation simulation radially includes 55\% of the 15 $M_{\odot}$ star, and the outer portion of the simulation contains a wave damping layer (indicated by the grey shading).
    (Right Panel) A zoomed in view of the radial velocity in the Wave Generation simulation at the same instant.
    The waves are not visible because their velocities are much smaller than the convective velocities.
    {An animated version of this figure is available in the supplemental materials.}
    }
    \label{fig:dynamics}
\end{figure*}

Our strategy for determining the photometric variability from gravity waves is analogous.
We run short simulations of \textit{wave generation} by convection and ``record'' the waves as they leave the convection zone.
We also run a fully self-consistent simulation of convection and \textit{wave propagation} in a truncated stellar model.
To test our strategy, we apply to our wave recording a filter describing the effects of wave propagation in the truncated model.
The surface variability of the self-consistent simulation matches the filtered wave signal.
Having verified our strategy, we then apply a different filter associated with the real star to determine the stellar photometric variability.

\section{Results}
\label{sec:results}

To predict the photometric variability caused by gravity waves in massive stars, we used Dedalus \referee{\citep{burns_etal_2020,vasil_etal_2019_tcalc,lecoanet_etal_2019_tcalc}}{} to run the first 3D simulations of core convection that simultaneously solve the fully compressible equations, produce realistically low-Mach-number flows, and include the full radial extent of the core convection zone, including $r=0$.
We study two types of simulations: (1) ``Wave Generation'' simulations where we record the waves \citep{couston_etal_2018}, and a ``Wave Propagation'' simulation where we test our method of applying a filter to the ``wave recording'' to mimic the effects of the environment  \citep{lecoanet_etal_2021}.
The Wave Generation simulations span twice the radial extent of the core convection zone (55\% of the star by radius in our fiducial model) and include a damping region in the outer $\sim$10\% of the simulation domain by radius to prevent wave reflections.
The Wave Propagation simulation spans 93\% of the star by radius (99.99925\% by mass) and is time-integrated for hundreds of convective overturn timescales to allow power in standing wave modes to saturate.
While both types of simulations include wave generation and propagation, these simulation designs allow us to isolate and study (1) the wave luminosity generated by the convective core, and (2) the surface amplitude of gravity waves that have propagated through the radiative envelope.
Most of our simulations are based on a 15 $M_{\odot}$ MESA \referee{\citep{mesa1, mesa2, mesa3, mesa4, mesa5, mesa6}}{} stellar model with LMC metallicity at the zero-age main sequence (ZAMS); this model is representative of a star with no near-surface convection \citep{jermyn_etal_2022_window}.

In Fig.~\ref{fig:dynamics}, we display temperature fluctuations for the Wave Propagation simulation (left panel) and radial velocity for a Wave Generation simulation (middle \& right panels).
Wave excitation can be visibly seen as an overdensity of gravity waves near the core convection zone, e.g., in the bottom right of the middle panel.
In the right panel, we zoom in on the core convection zone; the RMS velocity of the core convection is $5.8 \times 10^{4}$ cm/s, similar to the MLT prediction of $6.7 \times 10^{4}$ cm/s from the MESA model. The convection in our simulation is turbulent with an average Reynolds number of $9.5 \times 10^{3}$ and Mach number of $9.5 \times 10^{-4}$.
These Wave Propagation and Wave Generation simulations use spherical coordinates with respectively 512 and 1024 resolution elements across the convective core (see Supplemental Materials Sec.~{2.4}).

\begin{figure*}[t]
    \centering
    \includegraphics[width=\textwidth]{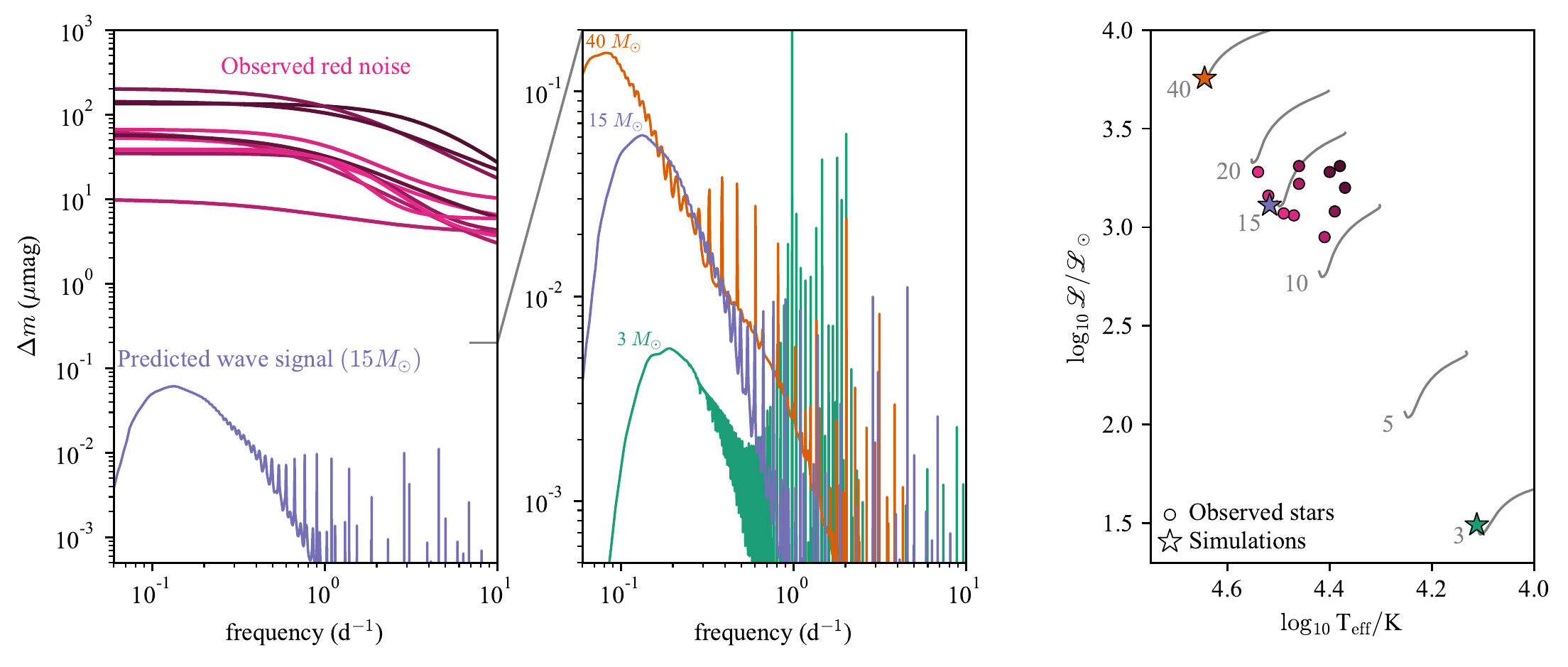}
    \caption{ 
    (Left panel) We show the predicted amplitude spectrum of gravity waves from our fiducial 15 $M_{\odot}$ simulation (purple line) and Lorentzian fits to red noise observations of stars with similar spectroscopic luminosity $\mathscr{L}$ and effective temperature $T_{\rm eff}$ \citep{bowman_etal_2020} (pink lines; the colors of the lines correspond to the colors of the points in the right panel).
    (Middle panel) A comparison of predicted amplitude spectra of gravity waves for the three stars that we simulate in this work.
    Note the different y-axis scaling between the left and middle panels.
    (Right panel) A spectroscopic HR diagram of the upper main sequence.
    The grey lines in the background show the main sequence evolution (from lower left to upper right) of nonrotating solar metallicity stars of [3, 5, 10, 15, 20, 40] $M_{\odot}$ from the MIST grids \citep{mist0,mist1}.
    The observed stars for which red noise fits are plotted in the left panel are shown as pink circles (darkened in color with distance from the MIST ZAMS in the $\log_{10} \mathscr{L} - \log_{10} T_{\rm eff}$ plane).
    The three simulated stars from the middle panel are displayed as star symbols.
    }
    \label{fig:observations}
\end{figure*}

The purpose of the Wave Propagation simulation is to test our method for applying a filter to the waves to mimic how the stellar environment modifies the waves \citep{lecoanet_etal_2019, lecoanet_etal_2021}.
To determine the wave amplitude in a star, we need (1) a recording of the waves generated by core convection and (2) a transfer function (``filter'') describing how waves propagate through the radiative envelope.
The wave luminosity spectrum is excited by stochastic, turbulent fluctuations in the convective core.
We measure (``record'') the wave luminosity of the traveling waves in Wave Generation simulations that have different resolutions and turbulent intensities. 
We find the wave luminosity is similar to theoretical predictions {\citep{lecoanet_quataert_2013}} and past simulations \citep{couston_etal_2018,lecoanet_etal_2021,lesaux_etal_2022}, and is independent of resolution provided there are at least 512 resolution elements across the convection zone (Supplemental Materials Sec.~{3.2}).
We separately evolve the Wave Propagation simulation for a very long time, and in doing so self-consistently evolve the generation and propagation of gravity waves in an environment similar to the star.
We build a transfer function (``filter'') using the gravity wave eigenvalues and eigenfunctions associated with the wave cavity of the Wave Propagation simulation (Supplemental Materials Sec.~5.2).
We synthesize a wave signal by convolving the wave luminosity from the Wave Generation simulation with this filter, and find good agreement between wave perturbations measured at the surface of the Wave Propagation simulation and our synthesized signal, verifying our method, see Supplemental Materials Sec.~{5.3}.

The Wave Propagation simulation is not identical to the star, because it has different diffusivities and a different wave cavity due to only including 93\% of the star, see Supplemental Materials Sec.~{2.6}.
Our Wave Propagation simulation has mode lifetimes of $\lesssim 10$ years, allowing us to capture both convection and wave propagation, which is not possible for real stars where mode lifetimes are $\sim 10^5$ years.
Having verified our method using the Wave Propagation simulation, we then create a \textit{separate}  ``full'' transfer function based on the true MESA stellar stratification and the gravity waves associated with the star's wave cavity, which we calculate using GYRE \referee{\citep{townsend_teitler_2013,townsend_goldstein_zweibel_2018}}{}, see Supplemental Materials Sec.~6.
We pass the wave luminosity signal from the Wave Generation simulation through this ``full'' transfer function to predict the photometric variability due to gravity waves.

\begin{figure*}
    \centering
    \includegraphics[width=\textwidth]{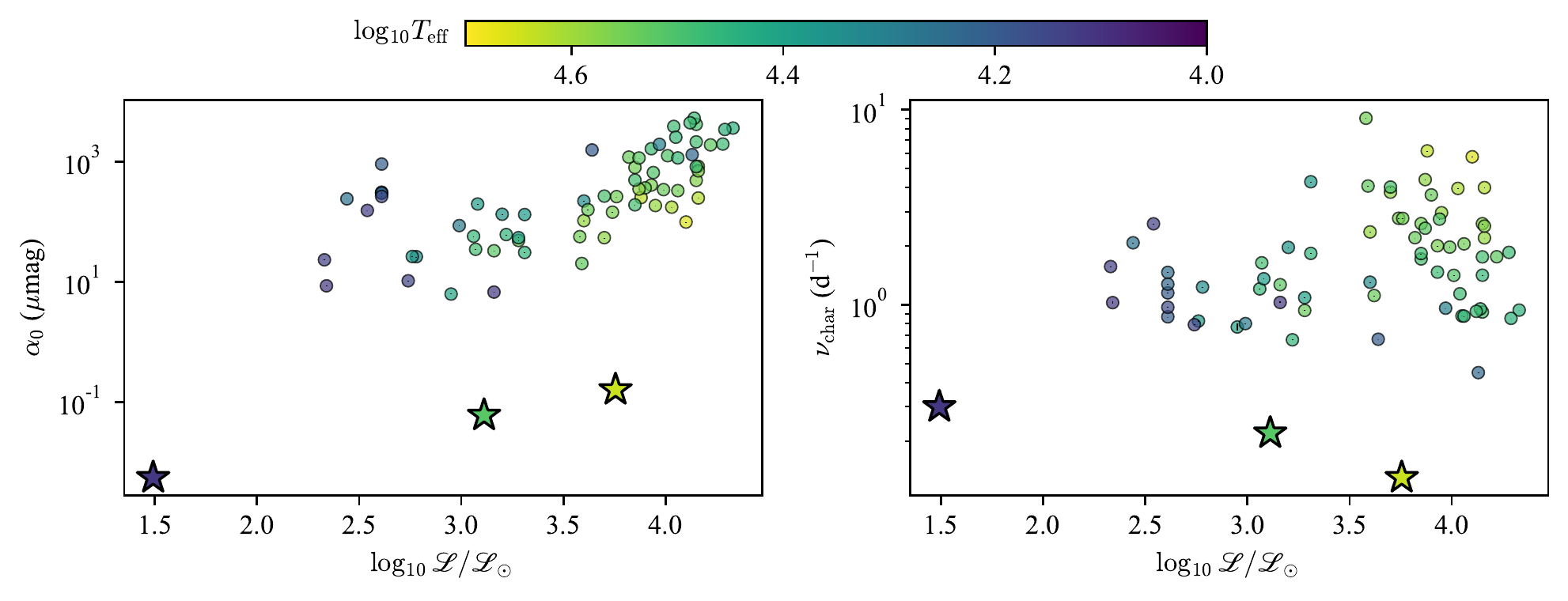}
    \caption{
    Shown are the magnitude $\alpha_0$ (left panel) and characteristic frequency $\nu_{\rm char}$ (right panel) obtained from fitting Eqn.~\ref{eqn:red_noise_fit} to our simulated gravity wave spectra (star symbols).
    We also include best-fits of red noise observations (circles) \citep{bowman_etal_2020}.
    All points are colored by $\log_{10} T_{\rm eff}$ and plotted against $\log_{10}\mathscr{L}/\mathscr{L}_{\odot}$.
    }
    \label{fig:ensemble}
\end{figure*}

In the left panel of Fig.~\ref{fig:observations}, we plot the predicted signal of gravity waves generated by core convection in our 15 $M_{\odot}$ star.
We make predictions for both the shape of the wave signal and its amplitude.
We find a wave signal characterized by a broad peak at a frequency near 0.1 d$^{-1}$.
The low frequency side of the peak is smooth and drops sharply, while the high frequency side of the peak decreases gradually and includes high amplitude, narrow peaks associated with resonant gravity modes.
The peak amplitude of the signal is roughly {${0.06\,\mu\text{mag}}$}.
We also display the publicly-available red noise observation fits \citep{bowman_etal_2020}, which are flat at low frequency, decrease sharply above $\sim$1 d$^{-1}$, and have amplitudes $\gtrsim 10$ $\mu$mag.
{
    We note that these fits underwent a pre-whitening procedure which iteratively removed the standing modes; the raw observations do include some peaks from standing modes.
    We include these fits primarily to provide a direct comparison between the typical \emph{amplitude} of red noise observations and our simulation-based predictions.
}

In the middle panel of Fig.~\ref{fig:observations}, we compare the simulated wave signals of three ZAMS stars of different masses.
These signals are similar, characterized by a broad peak at low frequency that has a sharp decrease on its low-frequency side, due to radiative damping, and a gradual decrease at high-frequency, due to less effective convective wave excitation and cancellation effects for high spherical harmonic degree (see Supplemental Materials Sec.~{6.2}).
The high frequency portions of the spectra are dominated by resonant wave peaks, and lower mass stars have more peaks and peaks of higher quality factor (see Supplemental Materials Figs.~8 \& 9). 
We find that the overall signal shifts to higher amplitude and lower frequency as the stellar mass increases. 
The stars considered in this work are displayed in a spectroscopic HR diagram in the right panel of Fig.~\ref{fig:observations}.
Our fiducial 15 $M_{\odot}$ LMC star is shown as a purple star, and we also simulate solar metallicity ZAMS stars of $3 M_{\odot}$ (green star) and $40 M_{\odot}$ (orange star).
The stars whose red noise signals we plotted are shown as pink circles, and they have $\log_{10}\mathscr{L}$ and $\log_{10} T_{\rm eff}$ which respectively differ by no more than $0.2$ from our fiducial star.

We fit our simulation spectra with Lorentzian profiles \citep{bowman_etal_2020} (Supplemental Materials Sec.~{6.3}),
\begin{equation}
    \alpha(\nu) = \frac{\alpha_0}{1 + \left(\frac{\nu}{\nu_{\rm char}}\right)^{\gamma}}.
    \label{eqn:red_noise_fit}
\end{equation}
In Fig.~\ref{fig:ensemble} we plot the amplitude $\alpha_0$ and characteristic frequency $\nu_{\rm char}$ of both the gravity wave signals from our simulations and observed photometric variability (red noise).
The amplitude $\alpha_0$ of the gravity wave signal increases in magnitude with increasing stellar luminosity and mass, similar to the amplitude of red noise.
The characteristic frequency $\nu_{\rm char}$ of the gravity wave signal decreases with increasing stellar luminosity and mass, whereas the characteristic frequency of red noise increases with increasing effective temperature but otherwise remains roughly constant as luminosity changes.

\section{Discussion}
\label{sec:discussion}

Our models predict surface signals of internal gravity waves generated by massive star core convection that are inconsistent with red noise observations. 
The photometric variability due to gravity waves is orders of magnitude lower than the observed red noise.
Our results suggest that a mechanism other than core-driven gravity waves must be responsible for the ubiquitous red noise signal.
{
    Our results are corroborated by recent 2D simulations with realistic luminosities \cite{lesaux_2023}, which found a wave luminosity spectrum with the same shape as ours, and which also found that wave propagation through the stellar envelope can be correctly modeled using linear theory.
}

While our simulations are the most realistic simulations of gravity wave generation by core convection to date, they do not include the following physical effects: magnetic fields, stellar models beyond the ZAMS, near surface convection zones, and rotation.
\begin{itemize}
    \item While the effects of magnetism are unclear, there are no currently known theoretical mechanisms by which they would substantially enhance wave generation by core convection.
    \item As stars age away from the ZAMS, they develop composition $\mu$-gradients outside of the core.
    These $\mu$-gradients modify the wave cavity and would affect the precise shape of the transfer function that we use to mimic the waves in the environment of a full star. 
    We do not expect composition gradients to have a large effect on the generation of waves.
    Therefore, while compositional gradients could affect the shape of the observed spectra,  there are no currently known theoretical mechanisms by which they would increase the amplitude of surface fluctuations.
    \item Our fiducial 15 $M_{\odot}$ LMC model is not expected to have near-surface convection, so our photometric variability predictions for this star could be directly compared to LMC data in this regime where near-surface convective zones are expected to be absent \citep{jermyn_etal_2022_window}.
    However, the observed galactic stars that we compare our results to are all expected to have near-surface convection, which could pollute or modify the wave signals we predict here.
    The 3 $M_{\odot}$ and 40 $M_{\odot}$ Wave Generation simulations we conducted are based on MESA models with solar metallicities and are expected to have near surface convection zones.
    Recent analyses of near-surface convection \citep{schultz_etal_2022} suggest that turbulence driven by subsurface convection could manifest as red noise, and future work should further examine the nature of turbulence generated by near-surface convection zones.
    \item Rotation affects the generation of waves, and could enhance the surface perturbations \citep{augustson_etal_2020}. 
    We performed one 15 $M_{\odot}$ Wave Generation simulation with a moderate rotation period $P_{\rm rot} = 10$ d (Supplemental Materials Sec.~7.2).
    We find that this rotation boosts the photometric signal amplitude to {$\alpha_0 = 0.21$ $\mu$mag}, still orders of magnitude weaker than the observed red noise.
    While this work cannot rule out the possibility that convection excites gravity waves to observable amplitudes in more rapidly rotating stars, the red noise signal is ubiquitous in both slow and rapid rotators, and there is no correlation between the stellar rotation rate and the red noise amplitude (see Supplementary Materials Sec.~7).
    Thus, the ubiquitous red noise signal cannot be due to gravity waves excited by core convection.
    {
    While rotational wave splitting would further modify the shape of the observed spectra, it would \textit{lower} the peak wave amplitude compared to our non-rotating predictions.
    }
\end{itemize}

While we have shown the photometric variability from convectively-excited gravity waves is not directly detectable, these waves may mix chemicals \citep{varghese_etal_2023} or transport angular momentum \citep{rogers_etal_2013}, leading to observable signals in more traditional $g$-mode asteroseismology \citep{decat_aerts_2002,pedersen_etal_2021,pedersen_2022}.
Even though the red noise signal is not the surface manifestation of gravity waves, it still carries valuable information about both the near-surface structure of massive stars as well as their masses and ages \citep{bowman_etal_2020, bowman_dornwallenstein_2022}.

\backmatter

\bmhead{Methods}

{
    All of the calculations in this work are based upon stellar models that we computed using Modules for Experiments in Stellar Astrophysics (MESA), version \texttt{r21.12.1} \citep{mesa1, mesa2, mesa3, mesa4, mesa5, mesa6}, see Supplemental Materials Sec.~1.
}

3D numerical simulations are performed using Dedalus \citep{burns_etal_2020} version 3; the specific version of the code run was obtained from the \texttt{master} branch of the Dedalus Github repository (\url{https://github.com/dedalusProject/dedalus}) at the commit with short-sha 
\texttt{29f3a59}.
We simulate the fully-compressible equations and assume the fluid is composed of a calorically perfect, uniform composition ideal gas.
We use a grid with high radial resolution from $r=0$ to {$r=1.1R_{\rm core}$, where $R_{\rm core}$ is the radius of the core convection zone,} and {we use} a lower radial resolution grid spanning the radiative envelope (See Supplemental Materials table 3).
The angular variation of all simulation variables are expanded using a basis of spin-weighted spherical harmonics; the radial variation of all simulations variables are expanded using a basis of radially-weighted Zernike polynomials in the convective core \citep{vasil_etal_2019_tcalc,lecoanet_etal_2019_tcalc}, and Chebyshev polynomials in the radiative envelope.
See Supplemental Materials, Sec.~2 for full simulation details.

{
    Most of our dynamical simulations are ``Wave Generation'' simulations, which extend from $r = 0$ to $2 R_{\rm core}$.
    These simulations include a damping layer near the outer boundary of the simulation.
    In these simulations, we measure the enthalpy and radial velocity perturbations on spherical shells at various radii throughout the simulation radiative zone at 30 minute intervals and use those measurements to calculate the wave luminosity as function of frequency and spherical harmonic degree.
    Wave Generation simulations are time-evolved for roughly 50 convective overturn timescales to build up sufficient statistics to measure the wave luminosity.
    We find that the wave luminosity is insensitive to the resolution and Reynolds number of calculations\referee{, suggesting that the wave luminosities that we report are similar to what would occur in a real star}{}.
    See Supplemental Materials, Sec.~3 for details on our measurements of the wave luminosity in Wave Generation simulations.
}

{
    In order to connect the convective driving to observable magnitude fluctuations at the stellar surface, we require a transfer function.
    We derive a transfer function closely following the procedure laid out in appendices A and B of Ref.~\citep{lecoanet_etal_2019} with a few modifications, see Sec.~4 of the Supplemental Materials.
    We note that the transfer function includes an $\mathcal{O}(1)$ amplitude correction factor $A_{\rm corr}$ which must be calibrated using 3D numerical simulations before it can be used to precisely predict the magnitude fluctuations at the surface of a star.
    In this work, we use $A_{\rm corr} = 0.4$.
}

{
    We run one ``Wave Propagation'' simulation, pictured in the left panel of Fig.~\ref{fig:dynamics}.
    This simulation includes 93\% by radius of the fiducial 15 $M_{\odot}$ stellar model, does not include a damping layer near the outer boundary, and is time-evolved for hundreds of convective overturn timescales in order to allow standing mode amplitudes to saturate.
    We evolve this simulation for 5 years, which is longer than the mode lifetime of almost all waves (the longest mode lifetime is $\sim$10 years, see Supplemental Materials Fig.~5).
    We measure the entropy perturbations at the outer boundary of this Wave Propagation simulation at 30 min intervals to determine the spectrum of gravity waves in a self-consistent, nonlinear simulation which includes most of the star.
    We verify that our prediction built from the transfer function and wave luminosity reproduces the measured surface perturbations well.
    We furthermore ran one more Wave Propagation simulation at a lower resolution (256 resolution elements across the convective core) and found that the same amplitude correction factor of $A_{\rm corr} = 0.4$ was required to align the transfer function prediction and surface perturbations, suggesting this factor is independent of the resolution, and thus applicable to stellar parameters.
    See Supplemental Materials, Sec.~5, for details on the surface spectrum of the Wave Propagation simulation, the verification of the transfer function, and the calibration of the amplitude correction factor.
}

{
    We predict the magnitude perturbation at the surface of a star from the nonadiabatic gravity wave eigenvalues and eigenvectors calculated using GYRE} version \texttt{7.0} \citep{townsend_teitler_2013,townsend_goldstein_zweibel_2018}.
{
We retrieve the \texttt{GYRE} g-modes with the standard \texttt{VACUUM} boundary conditions, which impose zero pressure perturbation at the stellar surface; we tested other boundary conditions formulations (\texttt{DZIEM}, \texttt{UNNO}, \texttt{JCD}) and found that our results were insensitive to this choice.
We compute the limb-darkened, disk-averaged differential flux functions from Eqns.~12-14 of Ref.~\citep{townsend2003}, accounting for an arbitrary observing angle as in Eqn.~8 of Ref.~\citep{townsend2002},} using MSG, version \texttt{1.1.2} \citep{townsend_lopez_2023}{, which synthesizes stellar spectra and convolves those spectra with an instrumental passband to account not just for intrinsic luminosity changes but also changes in the observed magnitude that a chosen telescope would see.} 
{We use} the OSTAR2002 grid in our $15 M_{\odot}$ and $40 M_{\odot}$ calculations \citep{lanz_hubeny_2003} and an extended version of the MSG demo grid for the $3 M_{\odot}$ star \citep{castelli_kurucz_2003} and {we select for our instrumental passband} the TESS passband from the SVO Filter Profile Service \citep{rodrigo_solano_2020}.
{
After synthesizing the differential flux functions, we calculate the magnitude perturbation eigenfunction associated with each eigenvalue using the \texttt{GYRE} solutions and Eqn.~11 of Ref.~\citep{townsend2002}.
We then use those magnitude eigenfunctions to generate a transfer function, which we use alongside the wave luminosity from Wave Generation simulations to predict the photometric variability of gravity waves.
}
{
    The stellar transfer functions, synthesized spectra, and red noise fits are discussed and shown in Supplemental Materials, Sec.~6.
}

{
    We additionally run one Wave Generation simulation with a rotation period of $P_{\rm rot} = 10$ d; we assume uniform, rigid rotation and include rotational effects by adding the Coriolis term to the momentum equation.
    We measure the wave luminosity in the same way as in the non-rotating simulation, and for simplicity we use the non-rotating transfer function.
    See Supplemental Materials, Sec.~7, for details of this simulation and a discussion of the importance of rotation.
}

All plots were produced using matplotlib \texttt{v3.5.2} \citep{Hunter:2007,mpl3.5.2}, and the numpy \texttt{v1.22.4} \citep{harris2020array}, scipy \texttt{v1.8.1} \citep{virtanen_etal_2020}, and astropy \texttt{v5.1} \citep{astropy:2013,astropy:2018,astropy:2022} packages were used in our simulation pre-processing and post-processing.
MESA profiles I/O was handled using \url{https://github.com/wmwolf/py_mesa_reader}.

\bmhead{Supplementary information}

We refer the reader to the supplementary materials, which contain a full description of the Dedalus simulation implementation, usage details of MESA, {details of the transfer function derivation, and figures showing wave luminosity spectra, wave power spectra at the surface of the Wave Propagation simulation, transfer functions, and rotating simulation results.}

\bmhead{Code Availability}

All Python scripts, MESA inlists, and GYRE inlists used in this work are in the git repository located at \url{https://github.com/evanhanders/gmode_variability_paper}\referee{, which will be released when this manuscript is accepted for publication}{}.
This repository is additionally backed up in a zenodo repository \referee{that will be released simultaneously}{} \citep{zenodo}.
Most of the underlying logic that implements the equations in Dedalus and generates transfer functions from Dedalus or GYRE outputs are separately located in the pip-installable github repository \url{https://github.com/evanhanders/compressible_stars}, version 0.1\referee{, which will also be released simultaneously}{}.

\bmhead{Data Availability}

The data used to create all of the figures in this paper, as well as the scripts used to generate the simulations and figures, are available online in this zenodo repository: \citep{zenodo}. \referee{ (which will be released upon acceptance for publication).}{}

{
\bmhead{Corresponding Author} 
Please address correspondence and requests for materials to Evan H.~Anders.
}

\bmhead{Acknowledgments}

We thank Conny Aerts, {Mathias Michielsen, Timothy Van Reeth, Cole Johnston,} Kyle Augustson, Meng Sun, Will Schultz, Lars Bildsten, Dominic Bowman, Falk Herwig, and Paul Woodward for helpful discussions which improved the work in this manuscript.
{We are grateful to the two anonymous referees for taking the time to review our work and for their valuable feedback which improved the clarity of this manuscript and the supplemental materials.}
%DL is supported in part by the US National Aeronautics and Space Administration (NASA) grant 80NSSC20K1280.
Computations were conducted with support by the NASA High End Computing Program through the NASA Advanced Supercomputing (NAS) Division at Ames Research Center on Pleiades with allocation GIDs s2276.
This work improved thanks to fruitful discussions at the ``Probes of Transport in Stars'' KITP program in Fall 2021.
%This research was supported in part by the National Science Foundation under grant no. NSF PHY-1748958.

\bmhead{Funding}

EHA was supported by a CIERA Postdoctoral Fellowship.
This work was supported in part by NASA HTMS grant 80NSSC20K1280, NASA SSW grant 80NSSC19K0026, and NASA OSTFL grant 80NSSC22K1738.
RHDT acknowledges support from NSF grants ACI-1663696, AST-1716436 and PHY-1748958, and NASA grant 80NSSC20K0515.
The Center for Computational Astrophysics at the Flatiron Institute is supported by the Simons Foundation.
This research was also supported in part by the National Science Foundation (NSF) under Grant Number NSF PHY-1748958.

{
\bmhead{Individual Contributions}
EHA, DL, and MC conceived and designed the experiments.
EHA performed the experiments.
EHA, DL, MC, EK, and BPB analyzed the data.
EHA, DL, KJB, BAH, RHDT, GMV, and ASJ contributed materials/analysis tools.
EHA, DL, MC, BAH, and BPB wrote the paper and supplemental materials.
}

\bibliography{biblio,mesa}% common bib file
%%\input article.bbl


\end{document}